\documentclass[traditabstract]{aa}
\usepackage{natbib}
\bibpunct{(}{)}{;}{a}{}{,} % to follow the A&A style
\usepackage{graphicx}

\begin{document}

\title{A Physically-Motivated Photometric Calibration of M Dwarf Metallicity}
\author{Kevin C. Schlaufman \and Gregory Laughlin}
\institute{Astronomy and Astrophysics Department, University of
California, Santa Cruz, CA 95064; kcs@ucolick.org and laugh@ucolick.org}
\date{Received 20 May 2010 / Accepted 11 June 2010}

\abstract{The location of M dwarfs in the $V-K_{s}$-$M_{Ks}$
color-magnitude diagram (CMD) has been shown to correlate with
metallicity.  We demonstrate that previous empirical photometric
calibrations of M dwarf metallicity exploiting this correlation
systematically underestimate or overestimate metallicity at the extremes
of their range.  We improve upon previous calibrations in three ways.
First, we use both a volume-limited and kinematically-matched sample
of F and G dwarfs from the Geneva-Copehnagen Survey (GCS) to infer the
mean metallicity of M dwarfs in the Solar Neighborhood.  Second, we use
theoretical models of  M dwarf interiors and atmospheres to determine
the effect of metallicity on M dwarfs in the $V-K_{s}$-$M_{Ks}$ CMD.
Third, though we use the GCS to infer the mean metallicity of M dwarfs
in the Solar Neighborhood, our final calibration is based purely on
high-resolution spectroscopy of FGK primaries with M dwarf companions as
well as the trigonometric parallaxes and apparent $V$- and $K_s$-band
magnitudes of those M dwarf companions.  As a result, our photometric
calibration explains an order of magnitude more of the variance in the
calibration sample than previous photometric calibrations.  We use
our calibration to non-parametrically quantify the significance of
the observation that M dwarfs that host exoplanets are preferentially
in a region of the $V-K_{s}$-$M_{Ks}$ plane populated by metal-rich M
dwarfs.  We find that the probability $p$ that planet-hosting M dwarfs
are distributed across the $V-K_{s}$-$M_{Ks}$ CMD in the same way as
field M dwarfs is $p = 0.06 \pm 0.008$.  Interestingly, the subsample
of M dwarfs that host Neptune and sub-Neptune mass planets may also
be preferentially located in the region of the $V-K_{s}$-$M_{Ks}$
plane populated by high-metallicity M dwarfs.  The probability of this
occurrence by chance is $p = 0.40 \pm 0.02$, and this observation hints
that low-mass planets may be more likely to be found around metal-rich M
dwarfs.  The confirmation of this hint would be in contrast to the result
obtained for FGK stars, where it appears that metal-rich and metal-poor
stars hosts Neptune-mass planets with approximately equal probability.
An increased rate of low-mass planet occurrence around metal-rich M
dwarfs would be a natural consequence of the core-accretion model of
planet formation.}


\titlerunning{M Dwarf Metallicity}
\authorrunning{Schlaufman \& Laughlin}
\maketitle

\section{Introduction}

\defcitealias{bon05a}{B05}
\defcitealias{joh09}{JA09}

The determination of metallicity for M dwarfs is a very difficult problem
\citep[e.g.][]{gus89}.  Their cool atmospheres permit the existence
of many molecules for which molecular opacities are currently poorly
constrained.  As a result, the estimation of the continuum level of a
spectrum is challenging, rendering line-based metallicity indicators
unreliable.  The poorly constrained molecular opacity data currently
available makes the determination of metallicity through spectral
synthesis difficult as well.  For those reasons, alternative methods
must be employed to estimate M dwarf metallicities.

The main sequence lifetimes of M dwarfs are longer than the Hubble time,
so they have not yet departed much from the zero-age main sequence.
Consequently, M dwarfs might be expected to form a two-parameter
sequence in mass and metallicity, suggesting that a two-color
broad-band photometric calibration might constrain their properties.
There have been several attempts to obtain the metallicity of M dwarfs
using their photometric properties, including two recent breakthroughs.
\citet{bon05a} -- \citetalias{bon05a} hereafter -- had the subtle insight
to realize that M dwarfs in binary or multiple systems should have
metallicities commensurate with the easily-measured metallicity of an
FGK primary in the system.  In that way, \citetalias{bon05a} identified
a calibration sample of M dwarfs with metallicities securely determined
in one of two ways: (1) high-resolution spectroscopy of an FGK companion
and (2) high-resolution spectroscopy of M dwarfs for which $T_{eff}$
and $\log{g}$ could be fixed with photometric data.  For the former,
the metallicity is very likely the same as the metallicity inferred
from high-resolution spectroscopy of its FGK companion.  For the latter,
spectral synthesis after fixing $T_{eff}$ and $\log{g}$ with photometric
data eliminates some degeneracy and produces a reasonable metallicity
estimate.  They noted that low-metallicity M dwarfs have blue $V-K_{s}$
color at constant $K_{s}$-band absolute magnitude $M_{Ks}$, and they fit
a linear model to their calibration sample using $V-K_{s}$ and $M_{Ks}$
to predict [Fe/H].  \citet{joh09} -- \citetalias{joh09} hereafter
-- addressed the relative lack of high-metallicity M dwarfs in the
calibration sample of \citetalias{bon05a} and created an empirical model
in which the distance of an M dwarf above the field M dwarf main sequence
(MS) in the $V-K_{s}$-$M_{Ks}$ color-magnitude diagram (CMD) indicated
its metallicity.  The great insight of \citetalias{joh09} was that the
mean metallicity of a population of M dwarfs could be characterized by the
easily-measured mean metallicity of a similar population of FGK stars.
Indeed, they assumed that the field M dwarf MS was an isometallicity
contour with the same metallicity as a volume-limited sample of G and K
stars and fit a linear model using the distance above the field M dwarf
MS to predict [Fe/H].

M dwarfs are attractive targets around which to search for low-mass
planets because they have large reflex velocities and transit depths even
for low-mass and small-radius companions.  Given that the metallicity
of protoplanetary disks is a key parameter in models of planet formation
\citep[e.g.][]{lau04,ida04}, the metallicity of M dwarfs that host planets
will constrain the planet formation process in low-mass protoplanetary
disks.  Indeed, it is well-established that metal-rich FGK stars
are more likely to host giant planets \citep[e.g.][]{san04,fis05},
but there is also evidence to suggest that metal-rich FGK stars
are not much more likely to host Neptune-mass planets than their
low-metallicity counterparts \citep[e.g.][]{udr06,sou08,bou09}.  Already,
\citetalias{joh09} have used their model of M dwarf metallicity to suggest
that the M dwarfs that host planets are preferentially metal-rich.
However, \citetalias{joh09} did not address whether the apparent lack
of a correlation between FGK host stellar metallicity and the presence
of Neptune-mass planets extends to M dwarfs.

In this paper, we examine a calibration sample of M dwarfs with securely
estimated metallicities and we show that the models of \citetalias{bon05a}
and \citetalias{joh09} systematically underestimate or overestimate
metallicity at the extremes of the range of this calibration sample.
We demonstrate that a volume-limited and kinematically-matched sample
of Sun-like stars produces a better estimate of the mean M dwarf
metallicity in the Solar Neighborhood, and we use M dwarf models of
different metallicities from \citet{bar98} to improve on the technique
described in \citetalias{joh09}.  The position of an M dwarf in the
$V-K_{s}$-$M_{Ks}$ CMD remains an indicator of its metallicity, and
we use that fact to non-parametrically quantify the significance of
the observation that planet-hosting M dwarfs are preferentially in a
region of the $V-K_{s}$-$M_{Ks}$ plane populated by metal-rich M dwarfs.
Moreover, we identify for the first time a hint that the subsample of M
dwarfs that host Neptune and sub-Neptune mass planets may also be more
likely to be in the region of the $V-K_{s}$-$M_{Ks}$ CMD associated with
metal-rich M dwarfs.  We describe our analysis in \S2 and summarize our
findings in \S3.

\section{Analysis}

\subsection{Testing Previous Calibrations}

We first collect from \citetalias{bon05a} and \citetalias{joh09}
a calibration sample of M dwarfs in wide binary or multiple systems
with an FGK primary.  The metallicity of the FGK primary in the system
is straightforward to measure from a high-resolution spectrum, and if
the M dwarf secondary and the FGK primary formed in the same molecular
core, then the expectation is that the two should have commensurate
metallicities.  We collect 13 examples from \citetalias{bon05a}, selecting
only those M dwarfs with precise $V$-band magnitudes from CCD photometry.
We also collect six high-metallicity examples from \citetalias{joh09}.
We summarize this calibration sample of M dwarfs in the first 19 lines
of Table~\ref{tbl-1}.

\begin{figure*}
\resizebox{\hsize}{!}{\includegraphics{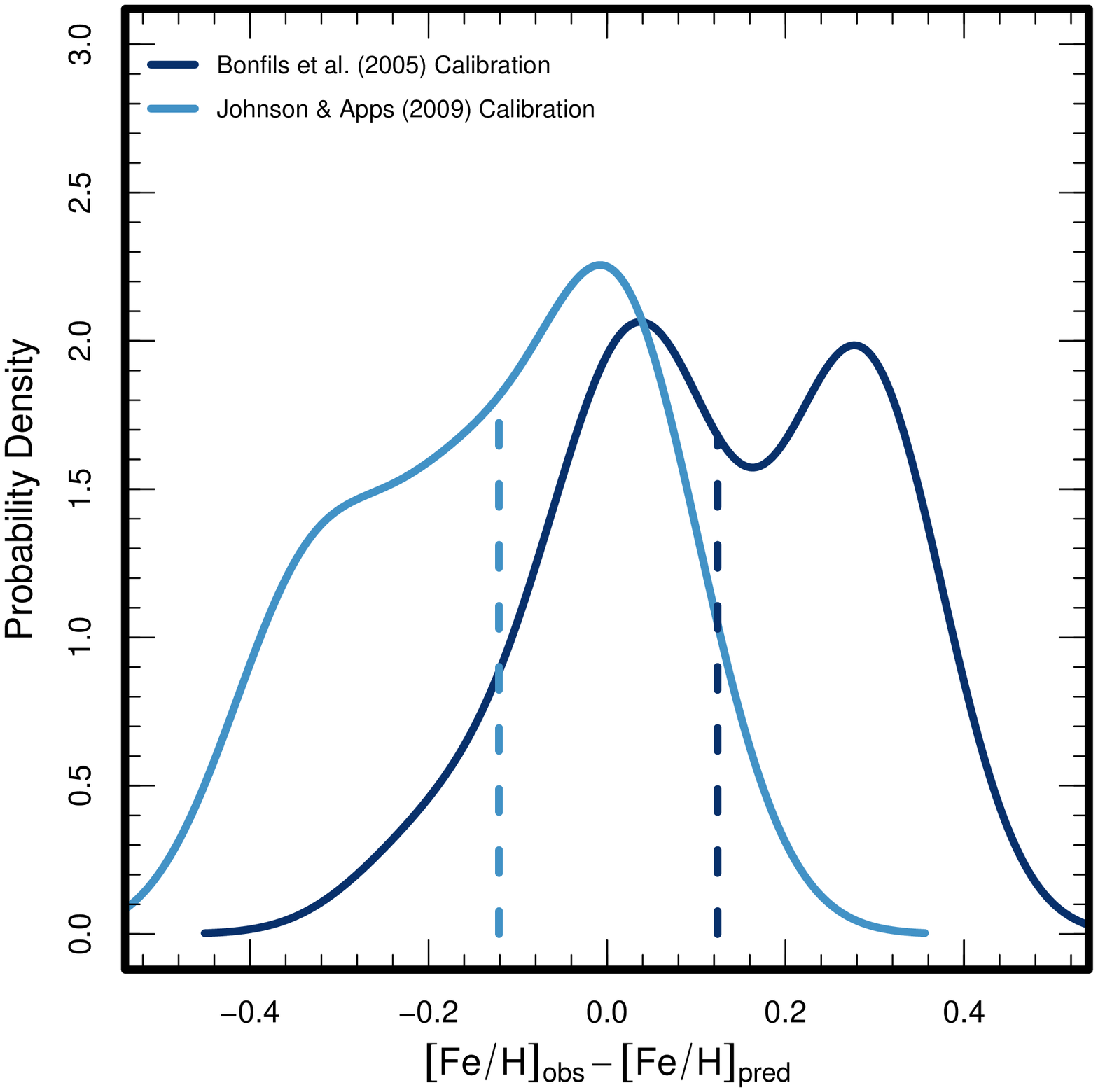}}
\caption{Optimally-smoothed residual distributions for \citet{bon05a} --
\citetalias{bon05a} hereafter -- and \citet{joh09} -- \citetalias{joh09}
hereafter.  In both cases the vertical dashed line indicates the
mean of the distribution.  The mean value of the \citetalias{bon05a}
residuals is 0.12 with standard deviation 0.16, while the mean value of
the \citetalias{joh09} residuals is -0.12 with standard deviation 0.12.
Note that the \citetalias{bon05a} distribution has a heavy tail at large
positive values (indicating systematically low [Fe/H] estimates) and the
\citetalias{joh09} distribution has a heavy tail at large negative values
(indicating systematically high [Fe/H] estimates).\label{fig01}}
\end{figure*}

We compute the metallicity predicted for this calibration sample from both
the \citetalias{bon05a} and \citetalias{joh09} relations and compare it
to the observed values.  Note that the \citetalias{bon05a} relation was
initially based on a calibration sample that included M dwarfs in binary
or multiple systems with FGK primaries (some with $V$-band magnitudes
deduced from photographic plates) as well as low-metallicity M dwarfs
with metallicity inferred from spectroscopy after fixing $T_{eff}$
and $\log{g}$ with photometric data.  Meanwhile, the \citetalias{joh09}
calibration sample included only the six metal-rich M dwarfs in binary
or multiple systems with FGK primaries listed in rows 14 through 19 in
Table~\ref{tbl-1}.  For those reasons, we believe that the 19 M dwarfs
with metallicities inferred from high-resolution spectroscopy of FGK
primaries in Table~\ref{tbl-1} is the largest and most reliable set of M
dwarf metallicities from which to verify previous calibrations.  We apply
both the \citetalias{bon05a} and \citetalias{joh09} relations to this
sample and compute the residual between each model and observation.  We
plot the distribution of residuals for both models in Figure~\ref{fig01},
and we find that the \citetalias{bon05a} relation systematically
underestimates  M dwarf metallicity and that the \citetalias{joh09}
relation systematically overestimates M dwarf metallicity.

\subsection{A Physically-Motivated Empirical Model of M Dwarf Metallicity}

As discussed in \S2.1 and Figure~\ref{fig01}, the models of
\citetalias{bon05a} and \citetalias{joh09} have non-negligible residuals
when applied to the calibration sample in Table~\ref{tbl-1}.  Still,
there is a correlation between the metallicity of an M dwarf and its
distance in the $V-K_{s}$-$M_{Ks}$ plane from the field M dwarf MS.
We attempted to improve the \citetalias{joh09} model by reassessing
both the zero point of the model and the direction from the M dwarf MS
in the $V-K_{s}$-$M_{Ks}$ plane best correlated with metallicity.

Recall that \citetalias{joh09} set the mean metallicity of the
Solar Neighborhood M dwarf sample equal to the mean metallicity of a
volume-limited sample of G0-K2 stars ($4.0 < M_{V} < 6.5$) from the SPOCS
catalog of \citet{val05}.  The SPOCS sample of \citetalias{joh09} was
based on a catalog of stars selected to have absorption lines deep enough
to enable high-precision radial velocity detection of exoplanets.  As a
result, the SPOCS sample is biased against metal-poor stars and therefore
potentially unsuitable for the determination of the average metallicity
in the Solar Neighborhood \citep[as noted in][]{val05}.  Alternatively,
the Geneva-Copenhagen Survey \citep[GCS -][]{nor04,hol07,hol09} of Solar
Neighborhood F and G dwarfs is magnitude-complete, kinematically-unbiased,
and free of the line depth bias inherent in the SPOCS catalog.  Though
the GCS metallicity estimates are based on Str{\"o}mgren $uvby\beta$
photometry and not high-resolution spectroscopy, the precision of the
GCS metallicities are sufficient when combined with the reduced bias of
the sample to provide a better estimate of the mean Solar Neighborhood
metallicity than the SPOCS sample.

\begin{figure*}
\resizebox{\hsize}{!}{\includegraphics{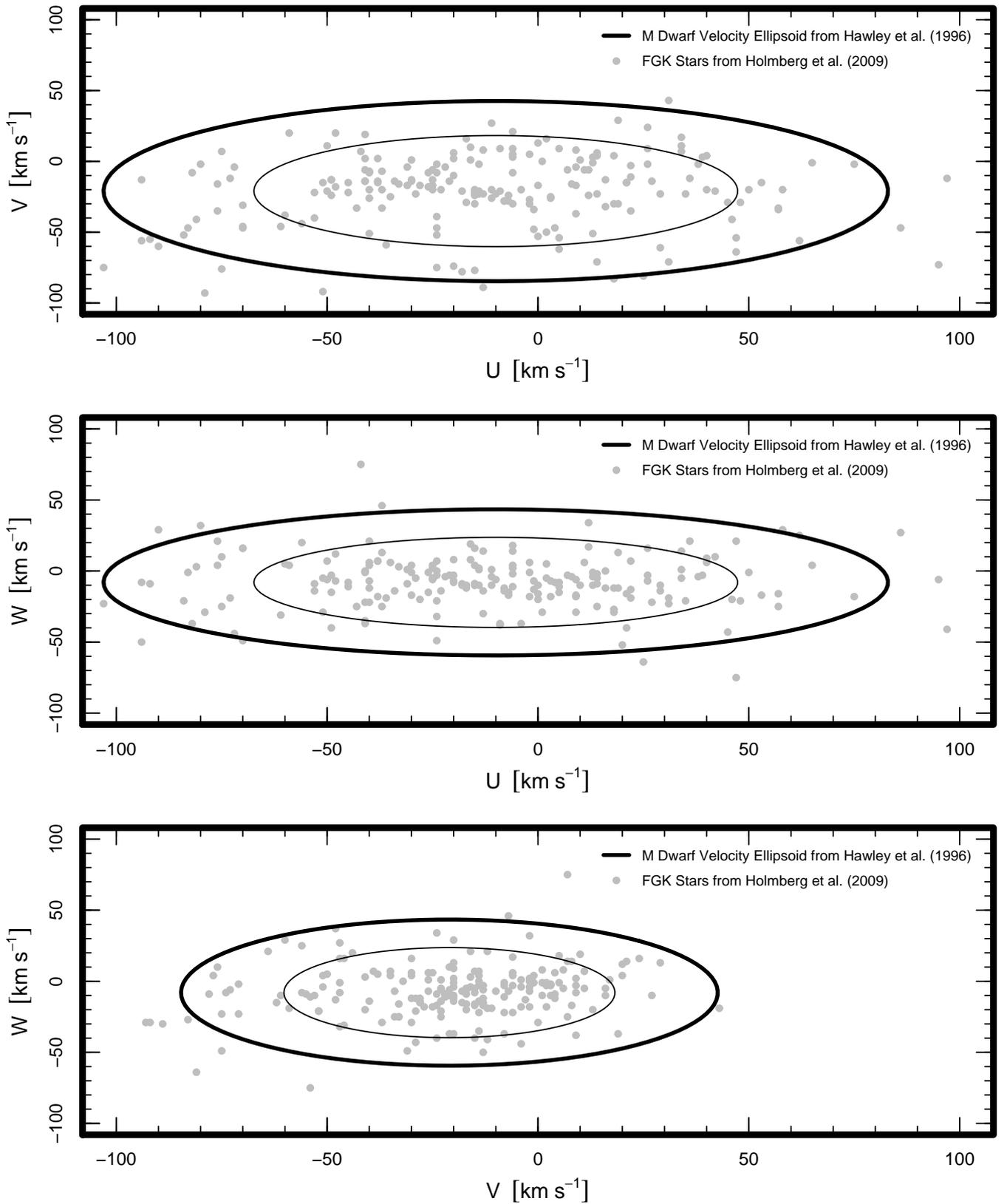}}
\caption{Velocity ellipsoids inferred for a volume-limited sample of
early M dwarfs from \citet{haw96} superimposed on the $UVW$ velocity
distribution of a volume-limited sample ($d< 20 \mbox{ pc}$) of Sun-like
stars from the Geneva-Copenhagen Survey \citep[gray points -][]{hol09}.
The light curve denotes the one-sigma region while the heavy curve
denotes the two-sigma region.  Bootstrap resampling of the \citet{hol09}
sample with the constraint that 68\% of each bootstrap sample lies within
the one-sigma contour and that the other 32\% lies within the two-sigma
contour produces a volume-limited and kinematically-matched population
metallicity of [Fe/H] $= -0.14 \pm 0.06$.\label{fig02}}
\end{figure*}

In addition, the $UVW$ kinematics of a volume-limited sample of M dwarfs
is not necessarily equivalent to the kinematics of a volume-limited
sample of FGK dwarfs.  Since the mean of a sample is sensitive to
outliers, and because Sun-like stars with outlier kinematics are
also likely to be outliers in metallicity, a kinematic-match is
important to determine the mean metallicity of Solar Neighborhood M
dwarfs in this way.  To address this point, we use the M dwarf $UVW$
distribution described by \citet{haw96} to create a volume-limited
and kinematically-matched sample of F and G dwarfs from \citet{hol09}
from which we infer the average metallicity of the Solar Neighborhood
M dwarf population.  In Figure~\ref{fig02} we superimpose the $UVW$
velocity-space distribution of local M dwarfs derived by \citet{haw96}
on top of the $UVW$ velocity-space distribution of F and G stars from
the GCS with parallax-based distance estimates that place them within
20 pc of the Sun.  We bootstrap resample from the subset of GCS stars
within 20 pc and with kinematics consistent with the M dwarf velocity
ellipsoid as defined in \citet{haw96}.  We ensure that 68\% of the GCS
stars in each bootstrap sample have $UVW$ velocities that place them
within the one-sigma contour of \citet{haw96} and that the rest of each
bootstrap sample lies within the two-sigma contour.  In the end, we find
that a volume-limited and kinematically-matched sample of F and G dwarfs
from the GCS survey has a mean metallicity of [Fe/H] $= -0.14 \pm 0.06$.
We obtain a similar result with a sample of GCS stars volume-limited in
the same way as the volume-limited SPOCS sample of \citetalias{joh09},
for which we find a mean Solar Neighborhood metallicity of [Fe/H] $=
-0.15 \pm 0.02$.  In this case, the superior statistics of the larger
volume-limited sample is enough to formally achieve a higher precision
than the volume-limited and kinematically-matched sample, though
the volume-limited sample is subject to a greater degree of possible
systematic error.  For that reason, we regard the volume-limited and
kinematically-matched result as likely more reliable.

The mean metallicity of our volume-limited and kinematically-matched GCS
sample suggests that the field M dwarf MS defined by \citetalias{joh09}
is an isometallicity contour with [Fe/H] $\approx -0.14$.  Note that
if the isometallicity contour corresponded to [Fe/H] $= -0.05$ as in
\citetalias{joh09}, five stars from Table~\ref{tbl-1} with [Fe/H] $<
-0.05$ would be to the right of the isometallicity contour indicating
[Fe/H] $> -0.05$.  Alternatively, if we assume that the isometallicity
contour corresponds to [Fe/H] $\approx -0.14$, then only two of the 19
stars are on the wrong side of contour.

\begin{figure*}
\resizebox{\hsize}{!}{\includegraphics{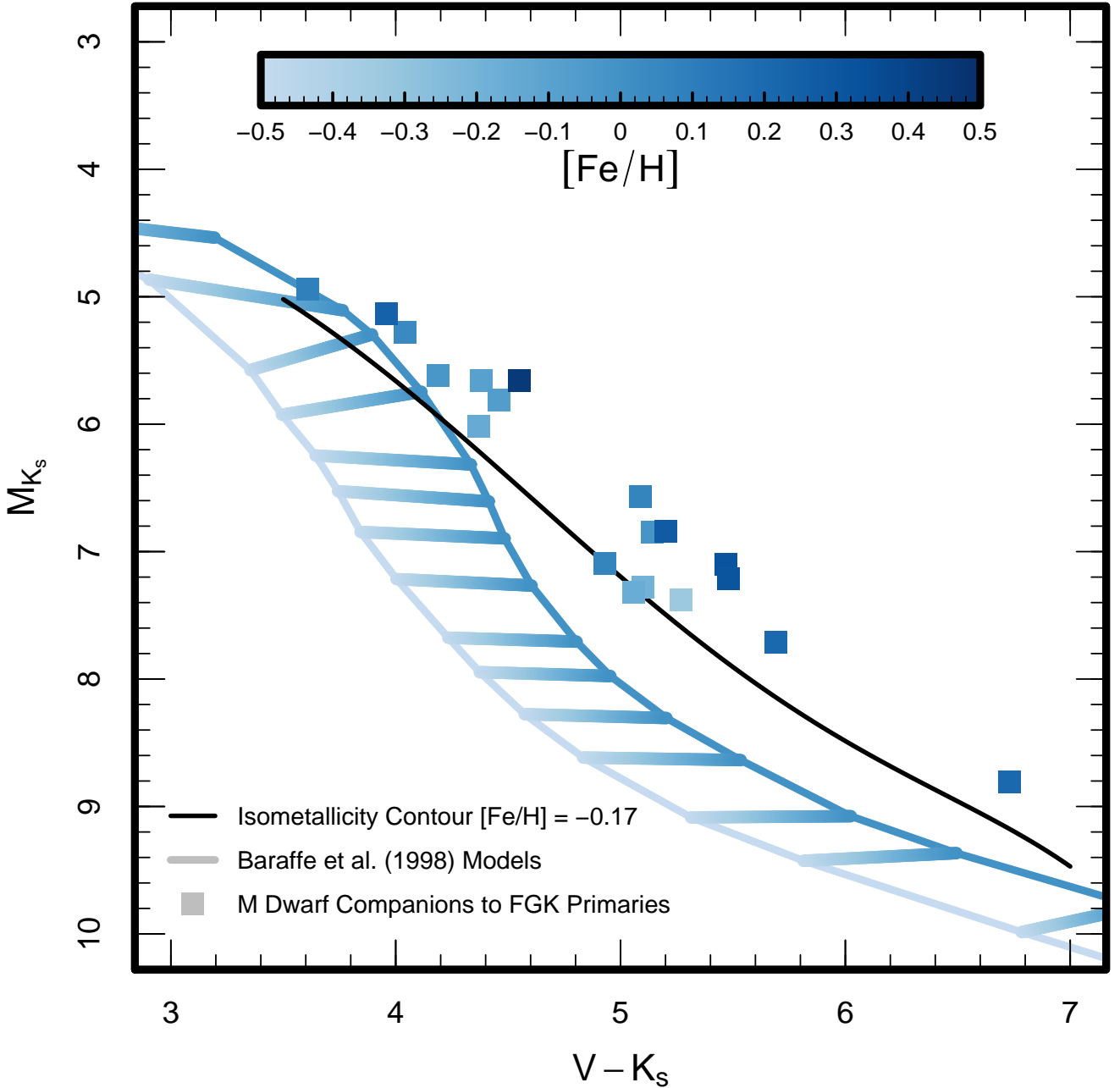}}
\caption{Position of M dwarfs with secure metallicities from
Table~\ref{tbl-1} (blue points) in the $V-K_{s}$-$M_{Ks}$ CMD in relation
to the field M dwarf MS from \citetalias{joh09} (black line) and the
theoretical isochrones of \citet{bar98} (blue lines).  The color of the
isochrone line gives its metallicity: [Fe/H] $= -0.5$ and $Y = 0.25$ on
the left and [Fe/H] $= 0$ and $Y = 0.275$ on the right.  Both isochrones
assume mixing-length parameter $l = 1$ for a 5 Gyr population, as there
is no detectable evolution in the $V-K_{s}$-$M_{Ks}$ CMD after after
about 3 Gyr.  The horizontal lines connect points of constant mass.
The models indicate that differences in metallicity should best correlate
with horizontal shifts in the $V-K_{s}$-$M_{Ks}$ plane.\label{fig03}}
\end{figure*}

We now determine which direction in the $V-K_{s}$-$M_{Ks}$ plane an
isochrone moves as a function of metallicity.  In Figure~\ref{fig03}
we plot the M dwarfs with securely determined metallicity from
Table~\ref{tbl-1}, along with the M dwarf MS from \citetalias{joh09} and
two different isochrones from \citet{bar98}.  We use the transformation
of \citet{car01} to transform the $K_{CIT}$ given in \citet{bar98}
into $K_{s}$.  The left-most isochrone corresponds to a population with
[Fe/H] $= -0.5$ and $Y = 0.25$ while the right-most isochrone corresponds
to a population with [Fe/H] $= 0$ and $Y = 0.275$.  Both isochrones
use mixing-length parameter $l = 1$ for a 5 Gyr population (there
is no detectable evolution in $V-K_{s}$-$M_{Ks}$ CMD after 3 Gyr).
The horizontal lines connect points of constant mass.  With all other
parameters constant, metallicity should best correlate with horizontal
shifts in the $V-K_{s}$-$M_{Ks}$ plane.  For that reason, we compute the
distance from the M dwarf MS in the horizontal direction for each M dwarf
with secure metallicity from Table~\ref{tbl-1}.  We then fit a linear
model using this distance as a predictor with [Fe/H] as the response.
We find that

\begin{eqnarray}\label{eq01}
\mbox{[Fe/H]} & = & 0.79 \Delta\left(V-K_{s}\right) - 0.17 \\
\Delta\left(V-K_{s}\right) & \equiv & \left(V-K_{s}\right)_{obs}-\left(V-K_{s}\right)_{iso} \nonumber
\end{eqnarray}

\noindent
is the optimal model.  In this case, $M_{Ks}$ as function of $V-K_{s}$
is given by the fifth-order polynomial with coefficients in increasing
order (-9.58933,17.3952,-8.88365,2.22598,0.258854,0.0113399)
from \citetalias{joh09}.  To aid in the calculation of
$\left(V-K_{s}\right)_{iso}$, we give the same curve with $V-K_{s}$ as
a function $M_{Ks}$: it is a fifth-order polynomial with coefficients in
increasing order (51.1413,-39.3756,12.2862,-1.83916,0.134266,-0.00382023).

We use two of the model selection criteria given in \citet{hoc76} to
evaluate all three models.  First, we compute the residual mean square
(RMS), defined as

\begin{eqnarray}\label{eq02}
\mbox{RMS}_{p} & = & \frac{\mbox{SSE}_{p}}{n-p}
\end{eqnarray}

\noindent
where $n$ is the number of data points, $p$ is the number of
predictors in the model, and $\mbox{SSE}_{p}$ is the residual
sum of squares for a $p$-term model.  In general, models with
smaller values of $\mbox{RMS}_{p}$ are best-suited to prediction.
For our model, we find that $\mbox{RMS}_{p} = 0.02$; for the model of
\citetalias{joh09} the value is $\mbox{RMS}_{p} = 0.04$ while for the
model of \citetalias{bon05a} the value is $\mbox{RMS}_{p} = 0.05$.  Next,
we compute the adjusted square of the multiple correlation coefficient
$R_{ap}^{2}$, which is widely used to judge the fit of a model.  A value
of $R_{ap}^{2} = 1$ indicates that a model explains all of the variance
in a sample, while $R_{ap}^{2} = 0$ indicates that the model explains
none of the variance.  $R_{ap}^{2}$ is defined as

\begin{eqnarray}\label{eq03}
R_{ap}^{2} & = & 1 - \left(n-1\right) \frac{\mbox{RMS}_{p}}{\mbox{SST}} \\
\mbox{SST} & \equiv & \sum{(y_i-\bar{y})^2} \nonumber
\end{eqnarray}

\noindent
where $y_i$ and $\bar{y}$ are the sample and its mean, respectively.
For our model, we find that $R_{ap}^{2} = 0.49$; for the model of
\citetalias{joh09} the value is $R_{ap}^{2} = 0.059$ while for the model
of \citetalias{bon05a} the value is $R_{ap}^{2} < 0.05$.  We note that
our model explains almost an order of magnitude more of the variance in
the calibration sample than either model presented in \citetalias{bon05a}
or \citetalias{joh09}.

Differing $T_{eff}$ scales have been well-noted as a source
of metallicity discrepancies in metallicity studies of the Solar
Neighborhood \citep[e.g.][]{hol07}.  The differing $T_{eff}$ calibrations
between the GCS and other surveys will not affect our results, as our
calibration (including the metallicity of the M dwarf MS) is based on
the horizontal distance in the $V-K_{s}$-$M_{Ks}$ CMD from the mean
M dwarf MS of \citetalias{joh09} of M dwarfs with metallicities known
from high-resolution spectroscopy of FGK primaries.  We only used the
GCS Str{\"o}mgren-based metallicities to establish the fact that the
mean Solar Neighborhood metallicity is closer to [Fe/H] $\approx -0.15$
than it is to [Fe/H] $\approx -0.05$ as argued by \citetalias{joh09}.
For that reason, our calibration is based purely on high-resolution
spectroscopy of FGK primaries as well as the trigonometric parallaxes
and apparent $V$- and $K_s$-band magnitudes of their M dwarf companions.

The M dwarfs in binary systems with FGK primaries that we use to fix our
calibration are not a volume-limited or kinematically-matched sample.
The volume-limit and kinematic-match were only necessary to verify the
fact that the mean metallicity of the M dwarf population in the Solar
Neighborhood is a well-defined quantity.  That verification is a necessary
precondition that must be established before any \citetalias{joh09}
style calibration using distance from the field M dwarf MS in the
$V-K_{s}$-$M_{Ks}$ CMD can even be considered.  Once the points along
the field M dwarf MS in the $V-K_{s}$-$M_{Ks}$ CMD are fixed to the
mean metallicity of the Solar Neighborhood, the metallicity of an M
dwarf with given $V-K_{s}$ color and absolute magnitude $M_{Ks}$ along
the curve is specified regardless of its position or velocity.  Indeed,
when we build our calibration using only the metallicities of M dwarfs
in binaries with FGK primaries, their $V$- and $K_{s}$-band magnitudes,
and trigonometric parallaxes, we find that the mean metallicity of the
Solar Neighborhood M dwarf MS based on the calibration sample ([Fe/H]
$= -0.17 \pm 0.07$) is statistically indistinguishable from the mean
metallicity inferred from the volume-limited and kinematically-matched
sample ([Fe/H] $= -0.14 \pm 0.06$).

\subsection{The Metallicity of M Dwarfs that Host Planets}

\begin{figure*}
\resizebox{\hsize}{!}{\includegraphics{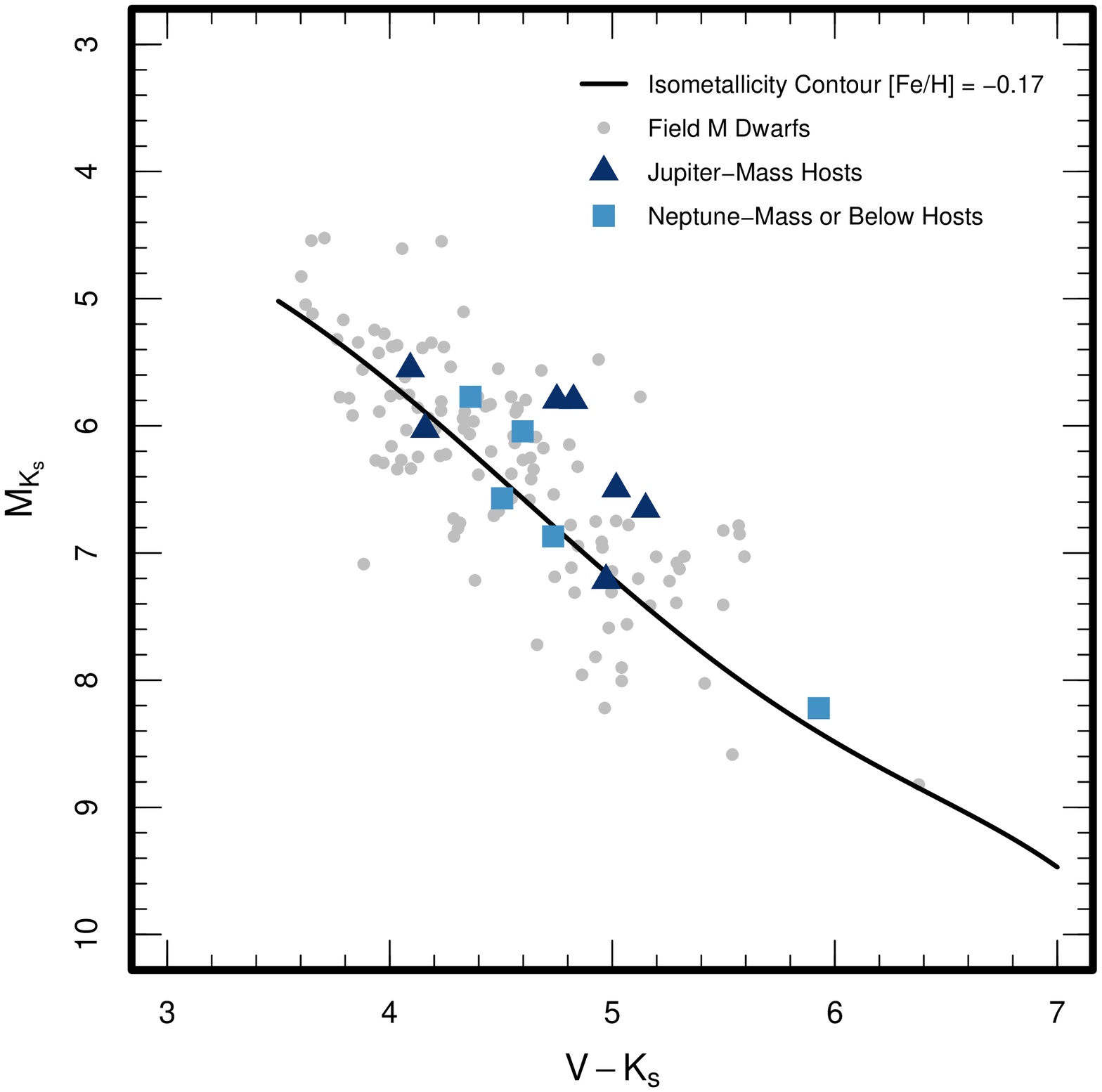}\includegraphics{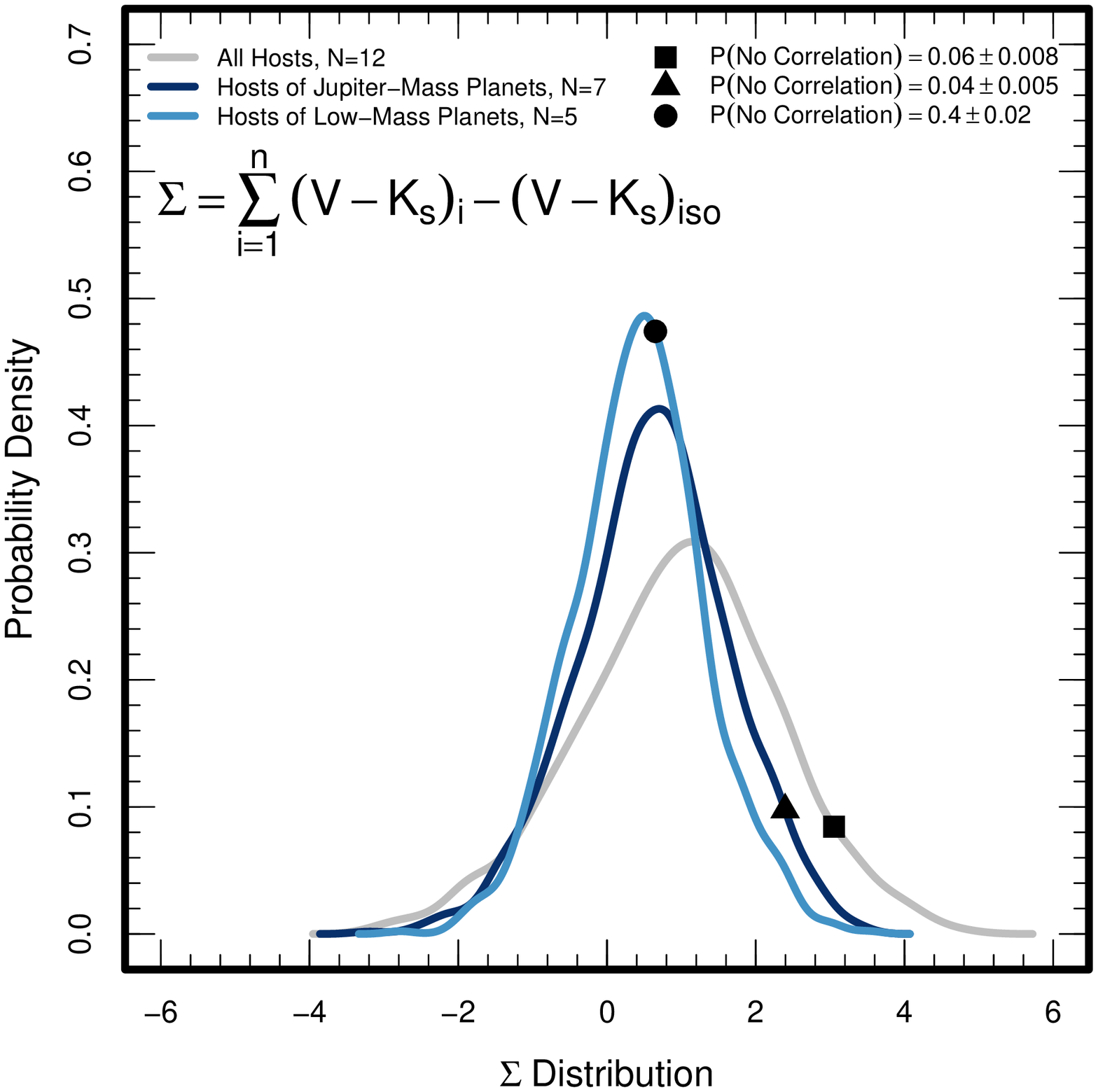}}
\caption{\emph{Left}: Position of M dwarfs known to host Jupiter-mass
planets (dark blue triangles) and Neptune-mass (or below) planets (blue
squares) in relation to a control sample of field M dwarfs (gray points)
and the field M dwarf MS from \citetalias{joh09} (black line).  Again,
like the high-metallicity M dwarfs, the M dwarfs that host planets
are concentrated to the right of the field M dwarf MS.  \emph{Right}:
Distribution of cumulative sample distances from the field M dwarf MS
of \citetalias{joh09}, which we assume to be a [Fe/H] $\approx -0.17$
isometallicity contour in the $V-K_{s}$-$M_{Ks}$ CMD.  Points to right
of the field M dwarf MS add their distance to the sum, while points
left of MS subtract their distance from the sum.  We generate each
distribution with a Monte Carlo simulation.  First, we randomly select
from the field M dwarf sample a number of stars equal to the number of
M dwarfs known to host planets of a certain type.  We then compute the
cumulative horizontal distance of that random subsample from the field M
dwarf MS.  We repeat this process 1000 times to generate the distribution
of sample cumulative horizontal distances from the field M dwarf MS given
no correlation between the presence of an exoplanet and the location of
its host in the $V-K_{s}$-$M_{Ks}$ CMD.  In all cases, we confirm the
findings of \citetalias{joh09} that the M dwarfs that host exoplanets
are preferentially to the right of the field M dwarf MS.  In particular,
we find that the probability $p$ that there is no correlation between
the location of an M dwarf in the $V-K_{s}$-$M_{Ks}$ CMD and its status
as an exoplanet host is $p = 0.06 \pm 0.008$.  For the subsample that
hosts Jupiter-mass planets, we find that the probability is $p = 0.04
\pm 0.005$.  More interestingly, we find that the probability that
there is no correlation between the location in the $V-K_{s}$-$M_{Ks}$
CMD and an M dwarf's status as the host of a Neptune-mass (or below)
exoplanet is $p = 0.40 \pm 0.02$.  If M dwarfs to the right of the field
M dwarf MS are metal-rich as suggested by the \citet{bar98} models and
argued by \citetalias{joh09}, then this observation may be evidence for
an increased incidence of low-mass planets around metal-rich low-mass
stars, a trend which is not observed in FGK stars.\label{fig04}}
\end{figure*}

We plot the location of M dwarfs that host planets in the left-hand
panel of Figure~\ref{fig04}.  The M dwarf models of \citet{bar98}
suggest that horizontal distance in the $V-K_{s}$-$M_{Ks}$ CMD best
correlates with metallicity.  To non-parametrically determine the degree
to which planet-hosting M dwarfs are preferentially found to the right
of the M dwarf MS, we need to quantify the likelihood that the cumulative
horizontal distance from the isometallicity contour of a randomly selected
sample of field M dwarfs can be as large as that observed in the sample
of M dwarfs that host planets simply by chance.

To address this issue, we create a control sample of field M dwarfs
selected from the Hipparcos \citep{van07} and Yale Parallax Catalogs
\citep{van95}.  We include in the control sample those M dwarfs from
the Hipparcos catalog that have parallaxes $\pi > 100$ mas precise to
better than 5\% and those M dwarfs from the Yale catalog that have
parallaxes $\pi > 100$ mas.  We use $V$-band photometry from each
catalog and we obtain $K_s$ photometry for both samples from the 2MASS
database \citep{skr06}.  For a sample of size $n$ we can compute the
statistic $\Sigma$:

\begin{eqnarray}\label{eq04}
\Sigma & = & \sum_{i=1}^{n} \left(V-K_{s}\right)_i-\left(V-K_{s}\right)_{iso}
\end{eqnarray}

\noindent
To characterize the likelihood that an observed value of $\Sigma$ for a
subsample with size $m$ of M dwarfs that host planets can be produced by
chance, we use a Monte Carlo simulation.  We randomly select a sample of
size $m$ from the 127 M dwarfs in the control sample, compute $\Sigma$
for that sample, save the result, and repeat the calculation 1000 times.
In that way, we can determine the distribution of $\Sigma$ expected under
the null hypothesis that M dwarfs that host planets are distributed
in the $V-K_{s}$-$M_{Ks}$ CMD in the same way as field M dwarfs.
We consider three sub-samples: (1) all planets hosts, (2) hosts of
Jupiter-mass planets, and (3) hosts of Neptune-mass (and below) planets.
We find that in case (1) $\Sigma = 3.43$ indicating only a $p = 0.06 \pm
0.008$ probability that the cumulative distance of the sample from the
isometallicity contour occurred by chance.  In case (2) we find $\Sigma =
2.39 \Rightarrow p = 0.04 \pm 0.005$ and in case (3) we find $\Sigma =
1.04 \Rightarrow p = 0.40 \pm 0.02$.  We summarize this calculation in
the right-hand panel of Figure~\ref{fig04}.

\subsection{Discussion}

The apparent position of planet-hosting M dwarfs in the region of the
$V-K_{s}$-$M_{Ks}$ CMD associated with known high-metallicity M dwarfs
tentatively suggests that metal-rich M dwarfs are more likely to host
Jupiter-mass and possibly Neptune-mass planets as well.  If this
correlation is confirmed in the future, it can be understood as a
natural consequence of the core-accretion model of planet formation
\citep[e.g.][]{lau04,ida04}.  Indeed, a more metal-rich protoplanetary
disk will almost certainly have a higher surface density of solids, and
that increased surface density enables the rapid formation of the several
Earth-mass cores necessary to accrete gas from the protoplanetary disk
before the gaseous disk is dissipated.  Moreover, it would be especially
interesting if the tentatively suggested correlation extends to the hosts
of Neptune-mass planets, as current evidence seems to suggest that the
probability that an FGK star hosts a Neptune-mass planet is not a strong
function of metallicity \citep[e.g.][]{udr06,sou08,bou09}.

If the tentatively suggested correlation between the presence of planets
and the metallicity of their host M dwarfs is eventually confirmed, it
might indicate a lower-limit on the amount of solid material necessary to
form planets.  To see why, recall that the mass of a protoplanetary disk
scales roughly as $M_{disk} \propto M_{\ast}$ and that the fraction of
solid material in a disk $f_{solid}$ scales roughly as $Z_{\ast}$ where
$Z_{\ast}$ is the metal content of the host star.  The total amount of
solid material in a protoplanetary disk will then scale like $M_{solid}
\propto f_{solid} M_{disk} \propto Z_{\ast} M_{\ast}$.  Minimum-mass Solar
Nebula models \citep[MMSN -][]{hay81} and observations of T Tauri disks
in star-forming regions suggest that protoplanetary disks around young
Solar-type stars are about 1\% the mass of their host stars, albeit with
significant scatter \citep[e.g.][]{har98}.  Combined with the fact that
the metal content of the Sun is $Z_{\odot} = 0.0176$ by mass, the total
solid mass in the MMSN was about $M_{solid} \approx 60~M_{\oplus}$.
This is a lower-limit, as more careful calculations suggest that the
protoplanetary disk around the Sun had $M_{solid} \sim 100~M_{\oplus}$
\citep[e.g.][]{lis93}.  In either case, this is a factor of a few to
ten greater than the $10~M_{\oplus}$ of material necessary to form the
core of gas or ice giant planet in the core-accretion model of planet
formation.  In the case of a Solar-metallicity mid-M dwarf with $M_{\ast}
= 0.3~M_{\odot}$, the total amount of solid material in the disk is 70\%
less, about $M_{solid} \approx 20~M_{\oplus}$.  This is factor of order
unity to a few times the mass necessary to form the core of a gas or ice
giant.  Since planet formation likely does not lock-up the entire solid
component of a protoplanetary disk in planets, reducing the total mass of
solids in the disk -- either by reducing the metallicity or mass -- will
also reduce the chances of forming a $10~M_{\oplus}$ core (and therefore
a gas or ice giant) before the parent protoplanetary disk is dissipated.

The confirmation of the hint of a correlation between the presence
of low-mass planets and M dwarf metallicity could be evidence of this
threshold solid mass necessary to form Neptune-mass planets.  Around FGK
stars, the same threshold solid mass suggests that a correlation between
the presence of low-mass planets and host star metallicity might occur at
one-third Solar metallicity, or [Fe/H] $= -0.5$.  This is just below the
typical metallicity of stars observed at high radial velocity precision
with HARPS \citep{sou08}.  This expected correlation might be verified
as larger samples of low-metallicity stars are surveyed at high radial
velocity precision or by transit surveys of nearby low-metallicity open
clusters (e.g. NGC 752 or IC 4756).

The hint of a correlation between the presence of Neptune-mass (and below)
planets and M dwarf metallicity tentatively suggests that searches for
low-mass planets around M dwarfs like the MEarth Project \citep{nut08}
could improve their yield by shading their target list toward M dwarfs
that have red $V-K_{s}$ colors at constant $K_{s}$-band absolute
magnitude $M_{Ks}$.  Note that since the absolute magnitude $M_{Ks}$
of an M dwarf depends only on the logarithm of its often poorly-known
distance, while $V-K_{s}$ depends linearly on its often poorly-known
$V$-band magnitude, the collection of high-quality CCD-based $V$-band
magnitudes for M dwarfs in the Solar Neighborhood could be the first
step towards maximizing the yield of planets around M dwarfs.

\section{Conclusion}

We showed that previous empirical photometric calibrations of M dwarf
metallicity systematically underestimate or overestimate metallicity
at the extremes of their range.  We derived a physically-motivated
model that explains an order of magnitude more of the variance in the
calibration sample that either the \citet{bon05a} or \citet{joh09} models.
We used the correlation underlying our model to non-parametrically
show that the probability $p$ that there is no relationship between
position of an M dwarf in the $V-K_{s}$-$M_{Ks}$ CMD and the presence
or absence of planets is $p = 0.06 \pm 0.008$.  For the subsample of
M dwarfs that host Jupiter-mass planets, the probability that there is
no correlation is $p = 0.04 \pm 0.005$.  Meanwhile, for the subsample
of M dwarfs that host Neptune-mass (or below) planets, we find that
the probability that there is no correlation is $p = 0.40 \pm 0.02$.
Since the models of \citet{bar98} suggest that the position of an M dwarf
in the $V-K_{s}$-$M_{Ks}$ CMD is a qualitative indicator of metallicity,
this observation tentatively suggests that metal-rich M dwarfs are
more likely to host planets and hints that the correlation may extend
to low-mass planets as well.  If this correlation is confirmed in the
future, it will be in contrast to planetary systems around FGK stars,
in which there appears to be only a weak connection between metallicity
and the presence of Neptune-mass planets.

\begin{acknowledgements}
We thank Connie Rockosi for useful comments and conversation and the
anonymous referee for many insightful suggestions that improved this
paper significantly.  This research has made use of NASA's Astrophysics
Data System Bibliographic Services, the Exoplanet Orbit Database and the
Exoplanet Data Explorer at exoplanets.org, and the NASA/IPAC Infrared
Science Archive, which is operated by the Jet Propulsion Laboratory,
California Institute of Technology, under contract with the National
Aeronautics and Space Administration.  This material is based upon
work supported under a National Science Foundation Graduate Research
Fellowship.
\end{acknowledgements}

\clearpage

\clearpage

\clearpage

\clearpage

\begin{table*}[h]
\centering
\caption[]{M Dwarfs in Binary Systems with an FGK Primary and Those
That Host Planets\label{tbl-1}}
\begin{tabular}{lccccccccc}
\hline\hline
System & $V$ & $K_{s}$ & $\pi$ & $V-K_{s}$ & $M_{Ks}$ &
Spectroscopic & Photometric & Comment & Reference \\
 & & [mag] & [mas] & [mag] & [mag] &
$\left[\mbox{Fe/H}\right]$ & $\left[\mbox{Fe/H}\right]$ & \\
\hline
Gl 105B & 11.67 & 6.57 & 138.72 & 5.1 & 7.28 & -0.19 & -0.14 & $\cdots$ & 1,2,3,4 \\
Gl 107B & 10.06 & 5.87 & 89.03 & 4.19 & 5.62 & -0.03 &  0.01 & $\cdots$ & 1,2,4,5 \\
Gl 166C & 11.17 & 5.9 & 198.08 & 5.27 & 7.38 & -0.33 & -0.05 & $\cdots$ & 1,2,4,5 \\
Gl 212 & 9.8 & 5.76 & 80.13 & 4.04 & 5.28 & 0.04 &  0.09 & $\cdots$ & 1,2,4,6 \\
Gl 231.1B & 13.42 & 8.28 & 51.76 & 5.14 & 6.85 & -0.02 &  0.12 & $\cdots$ & 1,2,4,5 \\
Gl 250B & 10.09 & 5.72 & 114.94 & 4.37 & 6.02 & -0.15 & -0.07 & $\cdots$ & 1,2,4,5 \\
Gl 297.2B & 11.8 & 7.42 & 44.47 & 4.38 & 5.66 & -0.09 &  0.13 & $\cdots$ & 1,2,3,4 \\
Gl 324B & 13.14 & 7.67 & 76.8 & 5.47 & 7.10 & 0.32 &  0.26 & $\cdots$ & 1,2,4,5 \\
Gl 53.1B & 13.6 & 8.67 & 48.2 & 4.93 & 7.09 & 0.07 & -0.17 & $\cdots$ & 1,2,4,5 \\
Gl 768.1B & 13.1 & 8.01 & 51.57 & 5.09 & 6.57 & 0.07 &  0.22 & $\cdots$ & 1,2,4,5 \\
Gl 783.2B & 13.94 & 8.88 & 48.83 & 5.06 & 7.32 & -0.16 & -0.19 & $\cdots$ & 1,2,4,5 \\
Gl 797B & 11.88 & 7.42 & 47.65 & 4.46 & 5.81 & -0.07 &  0.11 & $\cdots$ & 1,2,4,5 \\
Gl 81.1B & 11.21 & 7.6 & 29.43 & 3.61 & 4.94 & 0.09 & -0.08 & $\cdots$ & 1,2,4,5 \\
HD 46375B & 11.8 & 7.84 & 29.93 & 3.96 & 5.13 & 0.24 &  0.12 & $\cdots$ & 2,4,7 \\
HD 38529B & 13.35 & 8.8 & 23.57 & 4.55 & 5.66 & 0.45 &  0.27 & $\cdots$ & 2,4,7 \\
HD 18143C & 13.86 & 8.66 & 43.71 & 5.2 & 6.84 & 0.28 &  0.17 & $\cdots$ & 2,4,7 \\
55 Cnc B & 13.15 & 7.67 & 79.8 & 5.48 & 7.21 & 0.31 &  0.20 & $\cdots$ & 2,4,7 \\
HD 190360B & 14.4 & 8.71 & 62.92 & 5.69 & 7.71 & 0.21 &  0.10 & $\cdots$ & 2,4,7 \\
Proxima Cen & 11.11 & 4.38 & 772.33 & 6.73 & 8.81 & 0.21 &  0.14 & $\cdots$ & 2,4,7 \\
HIP 79431 & 11.34 & 6.589 & 69.46 & 4.751 & 5.80 & $\cdots$ &  0.35 & Jupiter & 2,8,9 \\
GJ 876 & 10.16 & 5.01 & 213.28 & 5.15 & 6.65 & $\cdots$ &  0.23 & 2 Jupiters + Super-Earth & 2,8,10 \\
GJ 317 & 12 & 7.028 & 109 & 4.972 & 7.22 & $\cdots$ & -0.20 & Jupiter & 2,11 \\
GJ 849 & 10.42 & 5.594 & 109.94 & 4.826 & 5.80 & $\cdots$ &  0.41 & Jupiter & 2,8,12 \\
GJ 179 & 11.96 & 6.942 & 81.38 & 5.018 & 6.49 & $\cdots$ &  0.20 & Jupiter & 2,8,13 \\
GJ 832 & 8.66 & 4.501 & 201.87 & 4.159 & 6.03 & $\cdots$ & -0.24 & Jupiter & 2,8,14 \\
GJ 649 & 9.7165 & 5.624 & 96.67 & 4.0925 & 5.55 & $\cdots$ & -0.03 & Saturn & 2,8,15 \\
GJ 436 & 10.67 & 6.073 & 98.61 & 4.597 & 6.04 & $\cdots$ &  0.10 & Neptune & 2,8,16 \\
GJ 581 & 10.57 & 5.836 & 160.91 & 4.734 & 6.87 & $\cdots$ & -0.22 & Neptune + 3 Super-Earths & 2,8,17 \\
GJ 674 & 9.36 & 4.855 & 220.24 & 4.505 & 6.57 & $\cdots$ & -0.24 & Neptune & 2,8,18 \\
GJ 176 & 9.97 & 5.607 & 107.83 & 4.363 & 5.77 & $\cdots$ &  0.06 & Neptune & 2,8,19 \\
GJ 1214 & 14.71 & 8.782 & 77.2 & 5.928 & 8.22 & $\cdots$ &  0.28 & Super-Earth & 2,20,21 \\
\hline
\end{tabular}
\tablebib{(1) \citet{bon05a}; (2) \citet{skr06}; (3) \citet{mer97};
(4) \citet{per97}; (5) \citet{gli91}; (6) \citet{egr92};
(7) \citet{joh09}; (8) \citet{van07}; (9) \citet{app10};
(10) \citet{mar98}; (11) \citet{joh07}; (12) \citet{but06};
(13) \citet{how10}; (14) \citet{bai09}; (15) \citet{joh10};
(16) \citet{man07}; (17) \citet{bon05b}; (18) \citet{bon07};
(19) \citet{for09}; (20) \citet{cha09}; (21) J. Irwin et al. (private communication)}
\end{table*}

\begin{thebibliography}{}
\bibitem[Apps et al.(2010)]{app10} Apps, K., et al.\ 2010, \pasp, 122, 156
\bibitem[Bailey et al.(2009)]{bai09} Bailey, J., Butler, R.~P., Tinney, C.~G.,
Jones, H.~R.~A., O'Toole, S., Carter, B.~D., \& Marcy, G.~W.\ 2009, \apj, 690,
743
\bibitem[Baraffe et al.(1998)]{bar98} Baraffe, I., Chabrier, G., Allard, F., \&
Hauschildt, P.~H.\ 1998, \aap, 337, 403
\bibitem[Bonfils et al.(2005a)]{bon05a} Bonfils, X., Delfosse, X., Udry, S.,
Santos, N.~C., Forveille, T., \& S{\'e}gransan, D.\ 2005, \aap, 442, 635
\bibitem[Bonfils et al.(2005b)]{bon05b} Bonfils, X., et al.\ 2005, \aap, 443,
L15
\bibitem[Bonfils et al.(2007)]{bon07} Bonfils, X., et al.\ 2007, \aap, 474, 293
\bibitem[Bouchy et al.(2009)]{bou09} Bouchy, F., et al.\ 2009, \aap, 496, 527
\bibitem[Butler et al.(2006)]{but06} Butler, R.~P., Johnson, J.~A.,
Marcy, G.~W., Wright, J.~T., Vogt, S.~S., \& Fischer, D.~A.\ 2006, \pasp, 118,
1685
\bibitem[Carpenter(2001)]{car01} Carpenter, J.~M.\ 2001, \aj, 121, 2851
\bibitem[Charbonneau et al.(2009)]{cha09} Charbonneau, D., et al.\ 2009, \nat,
462, 891
\bibitem[Egret et al.(1992)]{egr92} Egret, D., Didelon, P., McLean, B.~J.,
Russell, J.~L., \& Turon, C.\ 1992, \aap, 258, 217
\bibitem[Fischer \& Valenti(2005)]{fis05} Fischer, D.~A., \& Valenti, J.\ 2005,
\apj, 622, 1102
\bibitem[Forveille et al.(2009)]{for09} Forveille, T., et al.\ 2009, \aap, 493,
645
\bibitem[Gliese \& Jahrei{\ss}(1991)]{gli91} Gliese, W., \& Jahrei{\ss}, H.\
1991, Preliminary Version of the Third Catalogue of Nearby Stars, Tech. rep.
\bibitem[Gustafsson(1989)]{gus89} Gustafsson, B.\ 1989, \araa, 27, 701
\bibitem[Hawley et al.(1996)]{haw96} Hawley, S.~L., Gizis, J.~E., \&
Reid, I.~N.\ 1996, \aj, 112, 2799
\bibitem[Hartmann et al.(1998)]{har98} Hartmann, L., Calvet, N., Gullbring, E.,
\& D'Alessio, P.\ 1998, \apj, 495, 385
\bibitem[Hayashi(1981)]{hay81} Hayashi, C.\ 1981, Progress of Theoretical
Physics Supplement, 70, 35
\bibitem[Hocking(1976)]{hoc76} Hocking, R.~R.\ 1976, Biometrics, 32, 1
\bibitem[Holmberg et al.(2007)]{hol07} Holmberg, J., Nordstr{\"o}m, B., \&
Andersen, J.\ 2007, \aap, 475, 519
\bibitem[Holmberg et al.(2009)]{hol09} Holmberg, J., Nordstr{\"o}m, B., \&
Andersen, J.\ 2009, \aap, 501, 941
\bibitem[Howard et al.(2010)]{how10} Howard, A.~W., et al.\ 2010,
arXiv:1003.3488
\bibitem[Ida \& Lin(2004)]{ida04} Ida, S., \& Lin, D.~N.~C.\ 2004, \apj, 616,
567
\bibitem[Johnson et al.(2007)]{joh07} Johnson, J.~A., Butler, R.~P.,
Marcy, G.~W., Fischer, D.~A., Vogt, S.~S., Wright, J.~T., \& Peek, K.~M.~G.\
2007, \apj, 670, 833
\bibitem[Johnson \& Apps(2009)]{joh09} Johnson, J.~A., \& Apps, K.\ 2009, \apj,
699, 933
\bibitem[Johnson et al.(2010)]{joh10} Johnson, J.~A., et al.\ 2010, \pasp, 122,
149
\bibitem[Laughlin et al.(2004)]{lau04} Laughlin, G., Bodenheimer, P., \&
Adams, F.~C.\ 2004, \apjl, 612, L73
\bibitem[Lissauer(1993)]{lis93} Lissauer, J.~J.\ 1993, \araa, 31, 129
\bibitem[Maness et al.(2007)]{man07} Maness, H.~L., Marcy, G.~W., Ford, E.~B.,
Hauschildt, P.~H., Shreve, A.~T., Basri, G.~B., Butler, R.~P., \& Vogt, S.~S.\
2007, \pasp, 119, 90
\bibitem[Marcy et al.(1998)]{mar98} Marcy, G.~W., Butler, R.~P., Vogt, S.~S.,
Fischer, D., \& Lissauer, J.~J.\ 1998, \apjl, 505, L147
\bibitem[Mermilliod et al.(1997)]{mer97} Mermilliod, J.-C., Mermilliod, M., \&
Hauck, B.\ 1997, \aaps, 124, 349
\bibitem[Nordstr{\"o}m et al.(2004)]{nor04} Nordstr{\"o}m, B., et al.\ 2004,
\aap, 418, 989
\bibitem[Nutzman \& Charbonneau(2008)]{nut08} Nutzman, P., \& Charbonneau, D.\
2008, \pasp, 120, 317
\bibitem[Perryman et al.(1997)]{per97} Perryman, M.~A.~C., et al.\ 1997, \aap,
323, L49
\bibitem[Santos et al.(2004)]{san04} Santos, N.~C., Israelian, G., \&
Mayor, M.\ 2004, \aap, 415, 1153
\bibitem[Skrutskie et al.(2006)]{skr06} Skrutskie, M.~F., et al.\ 2006, \aj,
131, 1163
\bibitem[Sousa et al.(2008)]{sou08} Sousa, S.~G., et al.\ 2008, \aap, 487, 373
\bibitem[Udry et al.(2006)]{udr06} Udry, S., et al.\ 2006, \aap, 447, 361
\bibitem[van Altena et al.(1995)]{van95} van Altena, W.~F., Lee, J.~T.,
\& Hoffleit, E.~D.\ 1995, New Haven, CT: Yale University Observatory, |c1995,
4th ed., completely revised and enlarged,
\bibitem[Valenti \& Fischer(2005)]{val05} Valenti, J.~A., \& Fischer, D.~A.\
2005, \apjs, 159, 141
\bibitem[van Leeuwen(2007)]{van07} van Leeuwen, F.\ 2007, \aap, 474, 653
\end{thebibliography}
\end{document}